\documentclass[10pt,aps,prb,amssymb,amsmath,twocolumn,showpacs,floatfix]{revtex4-1}
\usepackage[utf8]{inputenc}
\usepackage{graphicx}
\usepackage{tikz}
\usepackage{pgf}
\usepackage{pgfplots}
\pgfplotsset{compat=newest}
\pgfplotsset{
        colormap={paraview_coolwarm}{
            rgb=(0.231373,0.298039,0.752941)
            rgb=(0.865003,0.865003,0.865003)
            rgb=(0.705882,0.0156863,0.14902)
        }
}
\usepgfplotslibrary{groupplots}
\usepackage{cool}
\usepackage[colorinlistoftodos,backgroundcolor=gray!10]{todonotes}
\usepackage{siunitx}
\usepackage{xspace}
\usepackage{url}
\usepackage{hyperref}
\usepackage{xcolor}
\usepackage[math]{blindtext}
\usepackage{gensymb} % for \degree

\newcommand*\chem[1]{\ensuremath{\mathrm{#1}}}

\newcommand*\mrm[1]{\ensuremath{\mathrm{#1}}}

\newcommand*\mbf[1]{\ensuremath{\boldsymbol{#1}}}

\newcommand*\Laplace{\mathop{{}\bigtriangleup}\nolimits}

\begin{document} %%%%%%%%%%%%%%%%%%%%%%%%%%%%%%%%%%%%%%%%%%%%%%%%%%%%%%%%%%
%------------------------------------------------------------------------------ 
% Title
%------------------------------------------------------------------------------
\title{Plasma flow around and charge distribution of a dust cluster in a rf discharge}

%------------------------------------------------------------------------------ 
% Authors
%------------------------------------------------------------------------------
\author{J. Schleede$^1$, L. Lewerentz$^{1}$, F. X. Bronold$^1$, R. Schneider$^{1}$, and H. Fehske$^1$}
\affiliation{$^1$Institut für Physik,
             Ernst-Moritz-Arndt-Universität Greifswald,
             D-17489 Greifswald,
             Germany}

%------------------------------------------------------------------------------ 
% Date
%------------------------------------------------------------------------------
\date{\today}

%------------------------------------------------------------------------------ 
% Abstract
%------------------------------------------------------------------------------
\begin{abstract}
We employ a particle-in-cell Monte Carlo collision/particle-particle particle-mesh (PIC-MCC/PPPM) simulation to study the plasma flow around and the charge distribution of a three-dimensional dust cluster in the sheath of a low-pressure rf argon discharge.
The geometry of the cluster and its position in the sheath are fixed to the experimental values, prohibiting a mechanical response of the cluster.
Electrically, however, the cluster and the plasma environment, mimicking 
also the experimental situation, are coupled self-consistently.
We find a broad distribution of the charges collected by the grains.
The ion flux shows on the scale of the Debye length strong focusing and shadowing inside and outside the cluster due to the attraction of the ions to the negatively charged grains whereas the electron flux is characterized on this scale only by a weak spatial modulation of its magnitude depending on the rf phase.
On the scale of the individual dust potentials, however, the electron flux deviates in the vicinity of the cluster strongly from the laminar flow associated with the plasma sheath.
It develops convection patterns to compensate for the depletion of electrons inside the dust cluster.
\end{abstract}
\maketitle

%\listoftodos

%------------------------------------------------------------------------------ 
% Introduction
%------------------------------------------------------------------------------
\section{Introduction}
\label{Sec:Introduction}

Fundamental to the physics of complex (dusty) plasmas is the charging of dust grains in a plasma. The dust attains huge negative electric charges due to the high mobility of the electrons in contrast to the ions.
The surrounding plasma shields the dust charges only partially so that they interact with each other via electrostatically screened Coulomb potentials.
These properties determine the electric forces between grains and result in complex self-organization phenomena like the formation of dust structures (e.g.\ Yukawa balls~\cite{arp_dust_2004,block_structural_2007}, dust clusters~\cite{ichiki_melting_2004}).
In most ground based laboratory experiments the dust grains are levitating above an electrode, where the electric field in the plasma sheath is finite and compensates for the gravitational force.
Thus, the grains are located in an environment of streaming ions and electrons.
In contrast to isotropic (bulk) plasmas the situation of streaming plasmas is  more intricate, leading to a broken symmetry of the charging currents towards the grain and asymmetric screening of the dust charge.
A dust grain in a flowing plasma acts like a lens for the plasma ions. 
This ion focusing effect gives rise to a wake behind the particle.
Inside the wake field there are points with enhanced positive space charge and increased potential~\cite{lampe_interactions_2000} that, in the case of multiple dust grains, will attract other negatively charged grains allowing the formation of aligned chains of dust grains along the direction of the flow \cite{bonitz_complex_2014}.

Single dust grain simulations have investigated in detail the charging of the dust grain \cite{matyash_finite_2006,ikkurthi_computation_2008,lapenta_simulation_1999}, wake potential \cite{lampe_interactions_2000,hou_induced_2003,hutchinson_nonlinear_2011,miloch_wake_2010} and the influence of the surrounding plasma properties like pressure \cite{hou_induced_2003,taccogna_dust_2012}, ion speed \cite{hutchinson_ion_2005,miloch_numerical_2007,lapenta_simulation_1999}, charge exchange \cite{taccogna_dust_2012,hutchinson_computation_2007,hutchinson_collisional_2013,rovagnati_effect_2007} and inhomogeneties in the plasma sheath \cite{hou_induced_2003,kompaneets_wakes_2014}.
In two-particle chains it has been found numerically that the negative charge of the downstream grain is reduced because the ion focus of the upstream 
grain increases the ion current onto the grain \cite{miloch_dust_2012,block_charging_2015}.

Miloshevsky et al.\ \cite{miloshevsky_self-confinement_2012,miloshevsky_dynamics_2012} investigated the charging of a rigid two-dimensional hexagonal dust cluster consisting of \num{19} dust grains and its impact on the plasma environment with a 2d PIC/MD simulation.
In absence of a streaming plasma (isotropic case) an energetically favorable configuration of the dust cluster was found for an inter-particle distance between dust grains of $\sim \num{0.6} \lambda_{D_e}$ due to the creation of a potential well by overlapping ion clouds around each dust grain.
In a streaming plasma with this minimal energy configuration the formation of a wake field with increased potential, ion and electron densities behind the cluster due to ion focusing effects was observed. 
Due to electron depletion within the cluster one grain obtained a positive charge.

Ikkurthi et al.\ \cite{ikkurthi_computation_2010} and Miloch et al. \cite{miloch_charging_2010,miloch_dust_2012,block_charging_2015} simulated three-dimensional arrangements of dust grains in flowing plasmas.
These works have examined aspects of parallel as well as perpendicular plasma flows through dust clusters.
They found that inner grains showed lower negative charge due to shadowing and boundary effects.
Even for 2d dust grain layers perpendicular to the flow the inner grains accumulated less charge.

However, most studies assume that dust particles are placed into a homogeneous background plasma with constant ion flow, while also neglecting ion-neutral collisional effects \cite{miloch_charging_2010,hutchinson_nonlinear_2011}.
The number of dust particles under investigation were mostly a few and their arrangements often in artificial simple geometry like chains \cite{miloch_dust_2012,miloch_wake_2010,hutchinson_forces_2011} or regular lattices \cite{miloshevsky_dynamics_2012,miloch_dust_2012,ikkurthi_computation_2010,matyash_pm_2010}.
The plasma sheath region where the dust grains are located has highly inhomogeneous ion and electron densities, a non-zero electric field, non-Maxwellian electron distribution \cite{godyak_measurement_1992,piel_dust_2015}, as well as a complex ion velocity distribution \cite{godyak_ion_1986-1} due to their acceleration in the sheath.
These properties of the plasma environments may have great influence on the charging of the individual dust particles inside a cluster.

\begin{figure*}[tbh]
\includegraphics[width=0.43\linewidth]{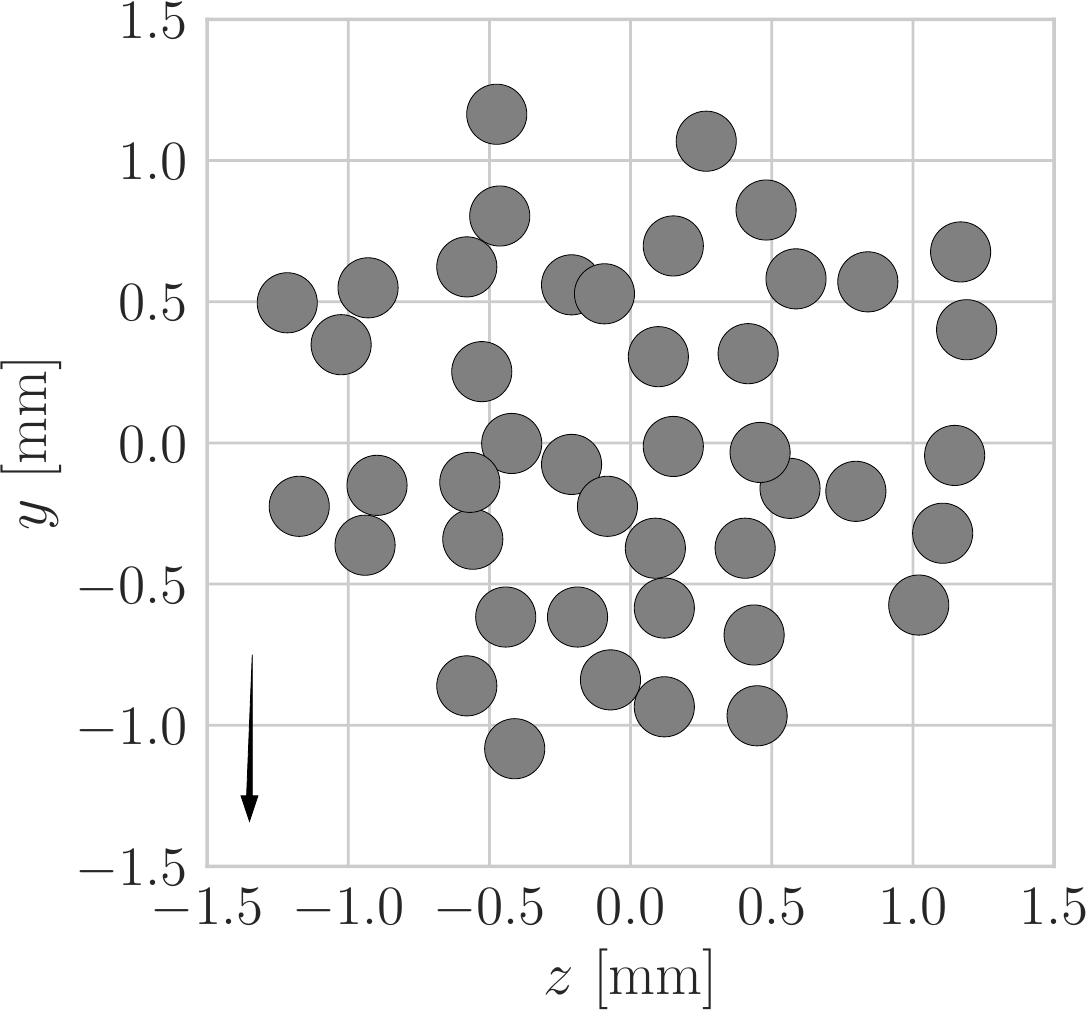}
\includegraphics[width=0.43\linewidth]{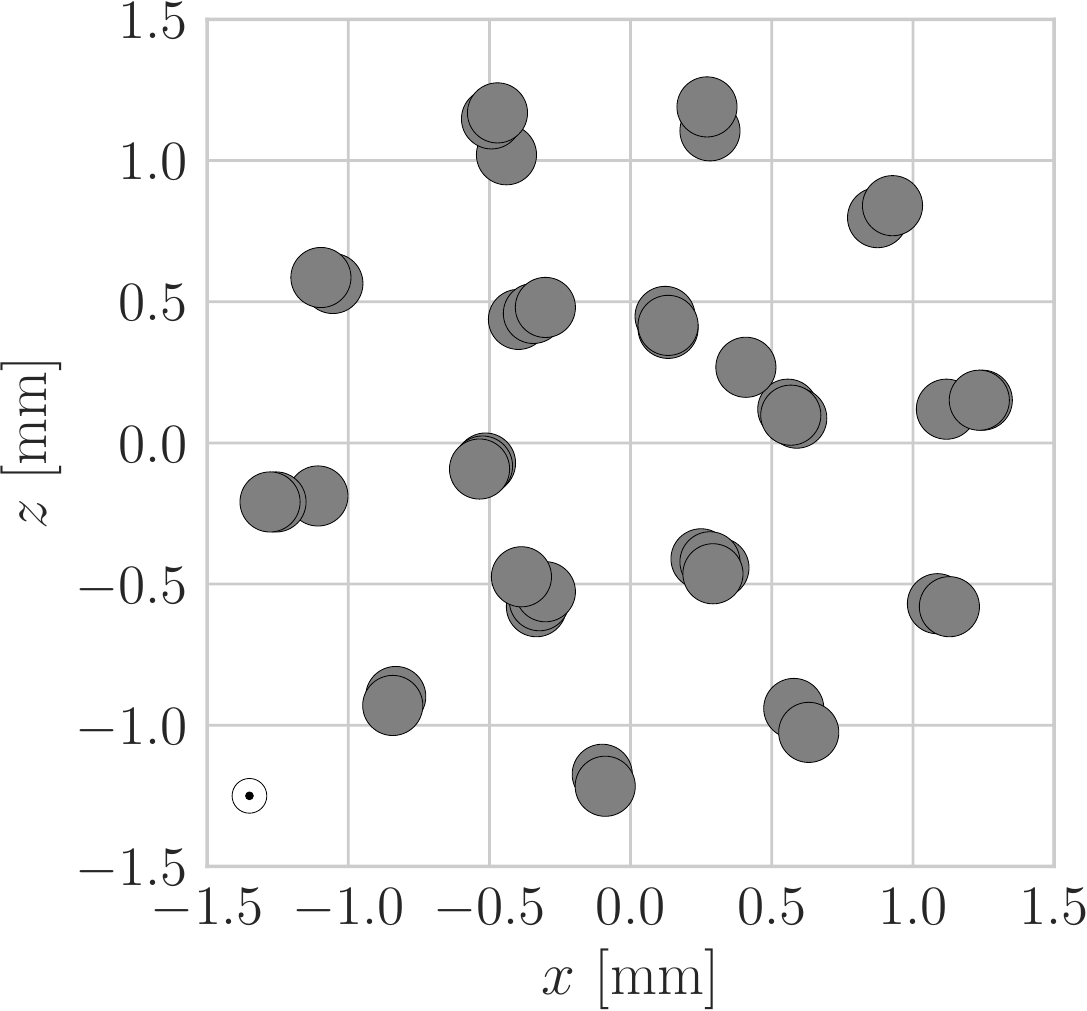}
\includegraphics[width=0.43\linewidth]{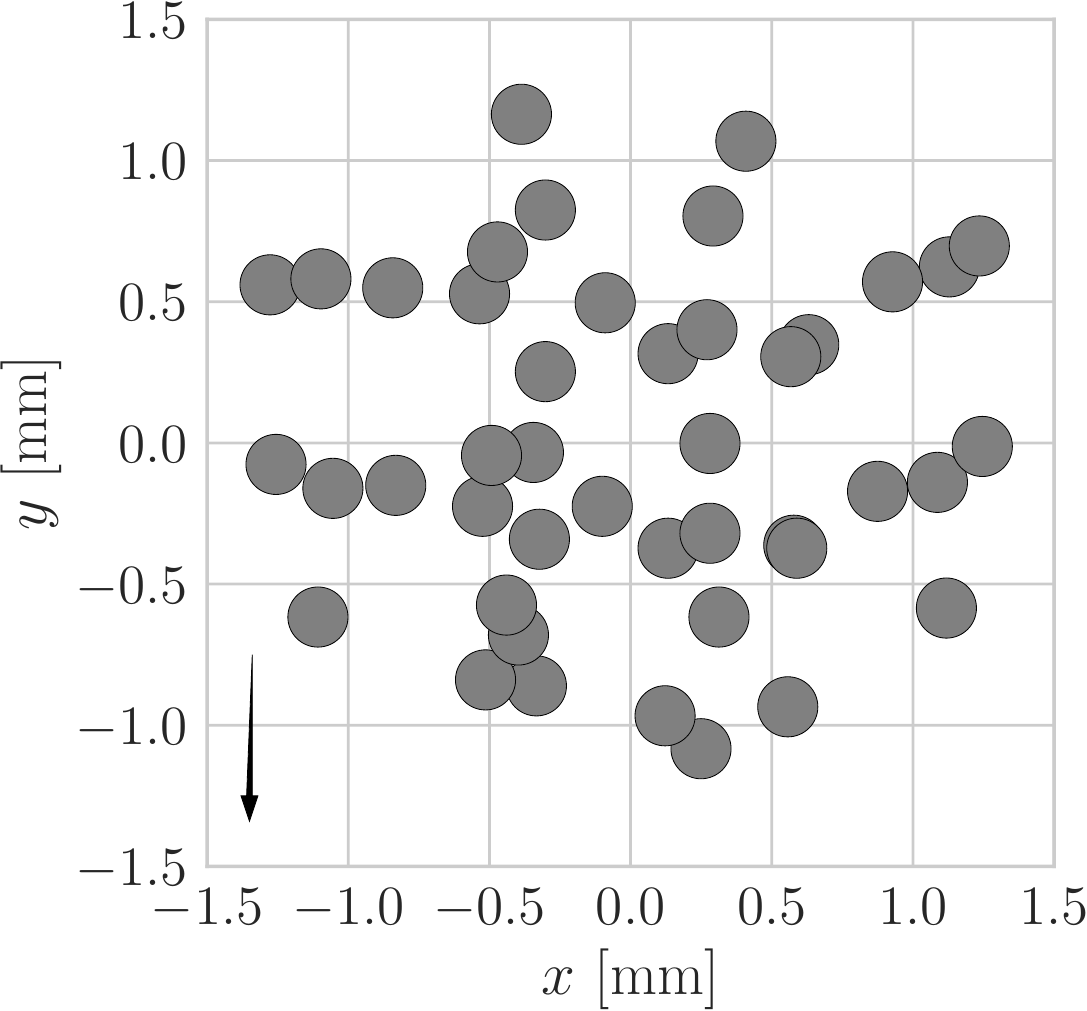}
\includegraphics[width=0.43\linewidth]{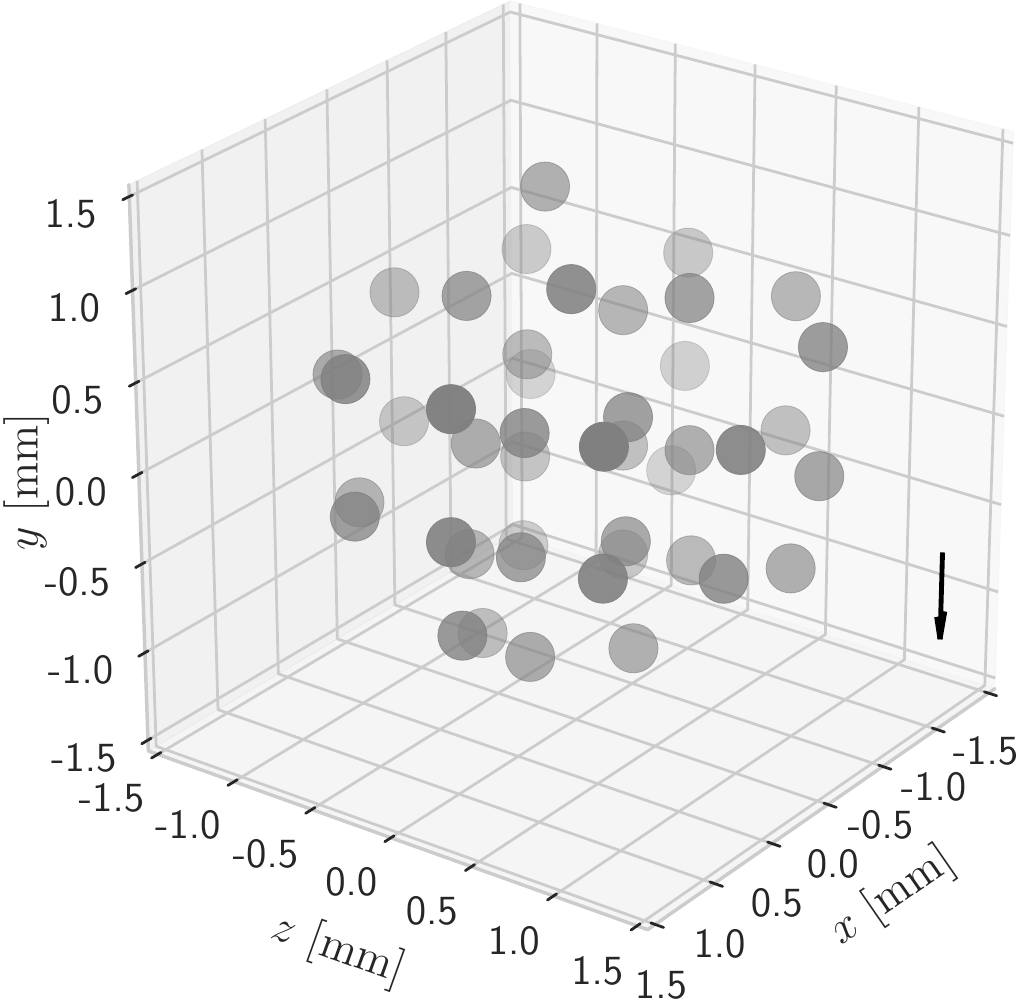}
\caption{\label{Fig:Dust}(Color online)
  Views on the positional data of the experimental dust cluster from different perspectives.
  The arrow indicates the net flow direction of the plasma ions through the cluster.
  Top left: projection along $x$ direction, top right: projection along $y$ direction, bottom left: projection along $z$ direction.
  Bottom right: 3d view, dust particles are shown with transparency to create the effect of depth.
  The cluster consists of $N_\mrm{D} = \num{44}$ particles with diameter $d_\mrm{D} = \SI{20}{\micro\meter}$ each.
  Particle sizes are increased for better visibility and the origin of the coordinate system is for this plot in the center of the cluster.}
\end{figure*}

In this work we perform a three-dimensional PIC-MCC/PPPM (Particle-in-Cell Monte Carlo collision/particle-particle particle-mesh) 
simulation of a low-pressure rf argon discharge containing a dust cluster of 44 particles in front of one of its electrodes. 
Considering the cluster as an electric probe, we focus on the plasma flow around and through it. In addition, we record its 
quasi-stationary charge distribution. The geometry of the cluster and its position in the sheath are assumed to be fixed. 
Mechanically, the grains are not allowed to react to the plasma. Only the electric response, that is, the charging-up of the 
particles and the modifications of the plasma flow associated with it, are investigated self-consistently.
The parameters of the simulations are mapped from experiment. Our simulation tracks thus the charge collected by the particles of the experimental cluster configuration in the plasma and visualises the plasma flow around it.
In addition, the dynamics of the rf discharge, especially acting on the electrons, is fully resolved.
% MUSS ANGEPASST WERDEN
On the scale of the Debye length we find strong shadowing and focusing of the ion flux due to the attraction of the ions to the negatively charged grains.
The electron flux develops convection cells inside the dust cluster to compensate for the electrons collected by the grains.
The grains are only able to charge-up to the values consistent with the surrounding plasma when the electron flux deviates on the scale below the Debye length from the unperturbed laminar flow associated with the sheath hosting the cluster. 
% BIS HIER

The paper is organized as follows. In Section \ref{Sec:Setup} we describe the experiment we simulate, giving the plasma parameters and the geometry of the dust cluster.
Section \ref{Sec:Method} explains the computer model and the details of our PIC-MCC/PPPM simulation scheme. The results are presented in Section \ref{Sec:Results}, showing data for the plasma background, the charging of the particles, and the plasma flows.
Concluding remarks are given in \ref{Sec:Conclusions}.

%------------------------------------------------------------------------------ 
% Setup
%------------------------------------------------------------------------------
\section{Experimental setup}
\label{Sec:Setup}

The dust cluster we want to model has been realized in an experiment by J. \ Schablinski, D. \ Block, and F. \ Greiner.
It consists of a parallel plate plasma reactor with horizontal circular electrodes of \SI{7}{\centi\meter} diameter and an adjustable vertical distance of \SIlist{3;5;7}{\centi\meter} between the lower and upper electrodes.
Located on the lower electrode, a glass box of \SI{20}{\milli\meter} side length that is open at its top and bottom was used for horizontal confinement of a cluster of dust particles~\cite{kroll_influence_2010,arp_confinement_2005} which is the main object of our investigation.
The plasma chamber, filled with argon gas at pressure $p = \SI{4.6}{\pascal}$ and temperature $T \approx \SI{300}{\kelvin}$, was driven by its lower electrode where a sinusoidal voltage with an amplitude of $U = \SI{100}{\volt}$ and a frequency of $f_\mrm{rf} = \SI{13.56}{MHz}$ was applied while the upper electrode was grounded.

The dust cluster consisted of $N_\mrm{D} = \num{44}$ hollow glass micro-spheres with a diameter of $d_\mrm{D} = \SI{20}{\micro\meter}$.
It had a total extent of roughly \SI{3 x 3 x 3}{\mm} and its center of mass was located at $y = \SI{5.6}{\mm}$ above the lower electrode.
Figure~\ref{Fig:Dust} shows the spatial structure of the dust cluster. The net flow of the plasma ions at the dust cluster location was directed towards the lower electrode at $y = \SI{0}{\centi\meter}$.
Resulting wake fields behind the dust particles created an attractive force onto nearby grains that led to vertically aligned dust particle chains along the $y$ axis.
This can be seen in the $xz$-projection by particles overlapping on top of each other (top right panel for Fig.~\ref{Fig:Dust}).
Furthermore the chains were arranged in two concentric rings due to electrostatic repulsion with a configuration of 6 inner and 10 outer chains.
The mean radius of the inner ring is about $r_\mathrm{inner}=\SI{0.55}{\mm}$ and $r_\mathrm{outer}=\SI{1.21}{\mm}$ for the outer ring.
The vertical chains do not consist of equal number of particles. The inner ring has more aligned particles than the outer one (left panels of Fig.~\ref{Fig:Dust}).
But it can be seen that distinctive layers existed in the radial ($xz$) direction where the particles are more or less situated at the same height.
The typical hexagonal structure for (two-dimensional) plasma crystals in the sheath could however not be found.
As far as the plasma environment of the cluster is concerned, it was noted that the discharge characteristics and the structure of the formed dust cluster did not change essentially for the different electrode distances.

In the simulation we assume the dust particles to have taken their equilibrium positions as determined by the experiment. Mechanically the particles do not respond to the plasma. We only describe the electric coupling leading to the charging of the particles and the modification of the plasma due to the charged particles.
The background plasma of the simulation mimicks at the position of the cluster the actual plasma conditions of the experiment although we do not include the glas box required for confining the particles. 
This is not necessary because in the simulation the cluster is assumed to be rigid. It acts thus like an extended electric probe. 
Thereby we can diagnose the charges of the grains collected in the experiment as well as the plasma flow around the dust particles present in the experiment.

%------------------------------------------------------------------------------ 
% Method of simulation
%------------------------------------------------------------------------------
\section{Method of simulation}
\label{Sec:Method}

The Particle-in-Cell (PIC) method allows the self-consistent description of the whole plasma experiment.
The real system consists of a huge number of particles, like electrons, ions and neutrals.
To make it possible to simulate such a system on a computer in reasonable time, super-particles are used instead of real particles.
Each super-particle represents a large number of real particles keeping the same charge-to-mass ratio $q/m$.
Because the equations of motion due (in general) to the Lorentz force only depend on the charge-to-mass ratio the trajectory of a super-particle is identical to a real particle.

All plasma particle species (electrons and argon ions) are treated fully kinetically. 
The neutrals are considered as a constant reservoir, calculated from the gas pressure used in the experiment. 
Therefore, a constant density distribution with a Maxwellian energy distribution of $T = \SI{300}{\kelvin}$ is realized by pseudo-particles. All the collisions of electrons and ions with the neutral gas particles are included by binary Monte-Carlo collisions, but the neutral gas particles are not affected by this.
A spatial grid mediates the long-range part of electrostatic force between particles.
Short-range interactions are mediated by Monte Carlo collisions (MCC)~\cite{brackbill_monte_1995,takizuka_binary_1977}.
Charge-exchange collisions between argon neutrals and ions, electron-impact ionization, elastic collisions, excitation and Coulomb collisions between charged particle species are taken into account.
Table \ref{Tab:Reactions} shows all the particle reactions incorporated into our simulation.

\begin{table}[bth]
\caption{\label{Tab:Reactions}
  Binary particle collisions and reactions used in the Monte Carlo collision (MCC) model~\cite{brackbill_monte_1995,takizuka_binary_1977} 
  for the argon plasma. Cross section data are taken from Phelps database (LxCat)
  ~\cite{phelps_application_1994,yamabe_measurement_1983}.
}
\begin{tabular}{l @{\quad} l}
\toprule
Name & Reaction \\
\colrule
Ionization & $\chem{e} + \chem{Ar} \rightarrow \chem{Ar^+} + 2\chem{e}$ \\
Charge exchange & $\chem{Ar} + \chem{Ar^+} \rightarrow \chem{Ar^+} + \chem{Ar}$ \\
Excitation & $\chem{e} + \chem{Ar} \rightarrow \chem{Ar^*} + \chem{e}$ \\
Coulomb collision & $\chem{e} + \chem{e}$, $\chem{e} + \chem{Ar^+}$, $\chem{Ar^+} + \chem{Ar^+}$ \\
Elastic collision & $\chem{e} + \chem{Ar}$, $\chem{Ar^+} + \chem{Ar}$ \\
\botrule
\end{tabular}
\end{table}

As in the standard PIC scheme the spatial resolution is on the order of the cell size, that is, on the order of half a Debye length in the plasma bulk (PIC scale), which for this simulation is $469\, \mu m$.
We use the Boris integration scheme\cite{Birdsall85} to propagate each PIC particle's position and velocity by discrete time steps $\Delta t$, hereby we only take into account the electric fields and neglect any magnetic fields:
\begin{eqnarray}
\mbf{x}_k &=& \mbf{x}_{k-1} + \mbf{v}_{k-\frac{1}{2}} \Delta t ~,\\
\mbf{v}_{k+\frac{1}{2}} &=& \mbf{v}_{k-\frac{1}{2}} + \frac{q}{m}\mbf{E}_k \Delta t  ~,
\end{eqnarray}
where $\mbf{x}_k$ is the particle position vector at time step $t_k = k \Delta t$, $\mbf{v}_{k+\frac{1}{2}}$ is the particle velocity vector at half integer time steps.
The electric field $\mbf{E}_k$ acting at a particle with charge-to-mass ratio $q/m$ is derived from the solution of Poisson's equation
\begin{eqnarray}
\Laplace \Phi_k &=& -\rho_k / \epsilon_0 ~,\\
\mbf{E}_k &=& -\nabla \Phi_k \quad .
\end{eqnarray}
The potential $\Phi_k$ and the charge density $\rho_k$ are mesh quantities located at the corners of each PIC cell.
The charge density is calculated with the standard PIC cloud-in-cell weighting scheme\cite{Birdsall85} using a box-like particle shape.
Given the charge density the discretized Poisson equation in finite difference form is solved by applying a simple and fast back-solve of an LU factorization.
The calculation of the LU factorization using the SuperLU library\cite{demmel_supernodal_1999,superlu_ug99} is only needed once per simulation run.
From the resulting potential the electric field for each PIC cell is calculated by finite differences.
The electric field acting at a particle is finally computed by using the same weighting scheme as for the charge density calculation.

In our special case the size of the dust particles is much smaller than the cell size, only about $1/10$.
In order to accurately detect the absorption of plasma particles at the small dust grains as well as to resolve the plasma particle motion in their vicinity, the PIC method is enhanced with a molecular dynamics (MD) approach.
In this kind of particle-particle particle-mesh (PPPM) model~\cite{matyash_finite_2006}, the long-range interaction of the dust grains with the charged plasma particles is treated by the standard PIC scheme.
Particles closer to the dust grains than a Debye length are handled by the MD approach.
This is due to the fact that the dust grains accumulate large values of charge so that the plasma in their vicinity is strongly modified and has large short-ranged density gradients.
Therefore, the interaction between dust grains and particles by the mean-field PIC scheme is not sufficient.
Instead, they interact directly via Coulomb forces.

The implementation follows Ikkurthi et al.~\cite{ikkurthi_computation_2010}: The PIC cell, where a dust grain is located, and its surrounding direct neighbor cells form a MD region.
Inside the MD region the electric field acting on the charged plasma species is calculated as
\begin{equation}
\mbf{E} = \mbf{E}_\mrm{PIC}^{\vphantom{\mrm{dust}}} - \mbf{E}_\mrm{PIC}^\mrm{dust} + \mbf{E}_\mrm{MD}^\mrm{dust} \quad.
\end{equation}
The part on the mean-field originating from the dust grains $\mbf{E}_\mrm{PIC}^\mrm{dust}$ is substituted by the exact Coulomb electric 
field 
\begin{equation}
\mbf{E}_\mrm{MD}^\mrm{dust} = \sum_{i=1}^{N_\mrm{D}} Q_i^\mrm{dust} \frac{\mbf{x}-\mbf{x}_i^\mrm{dust}}{|\mbf{x}-\mbf{x}_i^\mrm{dust}|^3}
\end{equation}
given by direct pairwise interaction between dust grains (charge $Q_i^\mrm{dust}$, position $\mbf{x}_i^\mrm{dust}$) and the plasma 
particle at position $\mbf{x}$.

To resolve the plasma particle motion on scales of the order of the dust grain size, it is necessary to reduce the time step inside the MD region.
We chose $\Delta t_\mrm{MD} = \Delta t/40$ to detect the absorption of particles onto the dust grains, otherwise fast particles (especially electrons) could make a positional move larger than the dust grain size, hereby literally jumping through a dust grain and missing the absorption event, which would lead to wrong dust charging values.

\begin{figure}[bt]
\includegraphics[width=0.99\linewidth]{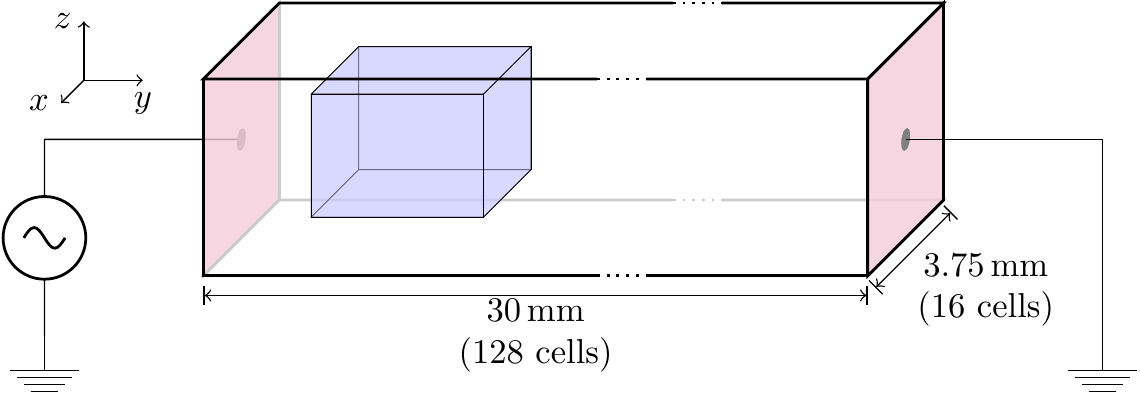}
\caption{(Color online)
  Schematic setup of the PIC plasma simulation.
  Note: the vertical $y$ direction is rotated by 90 degree clockwise. 
  The discharge is driven by an applied sinusoidal voltage between two metallic electrodes (red) along the $y$ direction.
  Periodic boundary conditions are applied along $x$ and $z$ direction.
  The blue box represents the MD region where the dust particles are located.
  Lengths are not drawn to scale.}
\label{Fig:Setup}
\end{figure}

In our simulation we make some geometric simplifications to reduce computational costs.
The plasma chamber is realized by a cuboidal domain in three-dimensional space, see Fig.~\ref{Fig:Setup}.
Metallic electrodes are located at both ends along the $y$ axis.
As our focus of interest is in the charging of the dust cluster and in the discharge characteristics in its vicinity it is not necessary to simulate the whole experimental volume of the plasma vessel.
The axial dimension along the discharge stays untouched ($L_y = \SI{3.0}{\cm}$), so that we can simulate a realistic plasma discharge as close as possible to the experimental conditions.
Because the dust cluster occupies only a tiny fraction of the experimental plasma volume and with the assumption of a homogeneous plasma (in space, not in time) across the lateral extent of the cluster, we shrink the lateral dimension to $L_x = L_z = \SI{0.375}{\cm}$ and apply periodic boundary conditions.
The simulation box provides enough space for the dust cluster itself.
Our computational domain of the rf discharge is represented by $\num{16 x 128 x 16}$ cubical PIC cells.
Each cell has an edge length of $\Delta x = \SI{234.28}{\micro\meter}$.
For the simulation of the argon plasma we use the same parameters for $T$, $p$, $f_\mrm{RF}$ and $U$ as given in Section \ref{Sec:Setup}.
Electrodes and dust particles are assumed to be perfect absorbers.

%------------------------------------------------------------------------------ 
% Results
%------------------------------------------------------------------------------
\section{Results}
\label{Sec:Results}

%------------------------------------------------------------------------------ 
% Results - Plasma Background
%------------------------------------------------------------------------------
\subsection{Plasma background}

\begin{figure}[tb]
\includegraphics[width=0.99\linewidth]{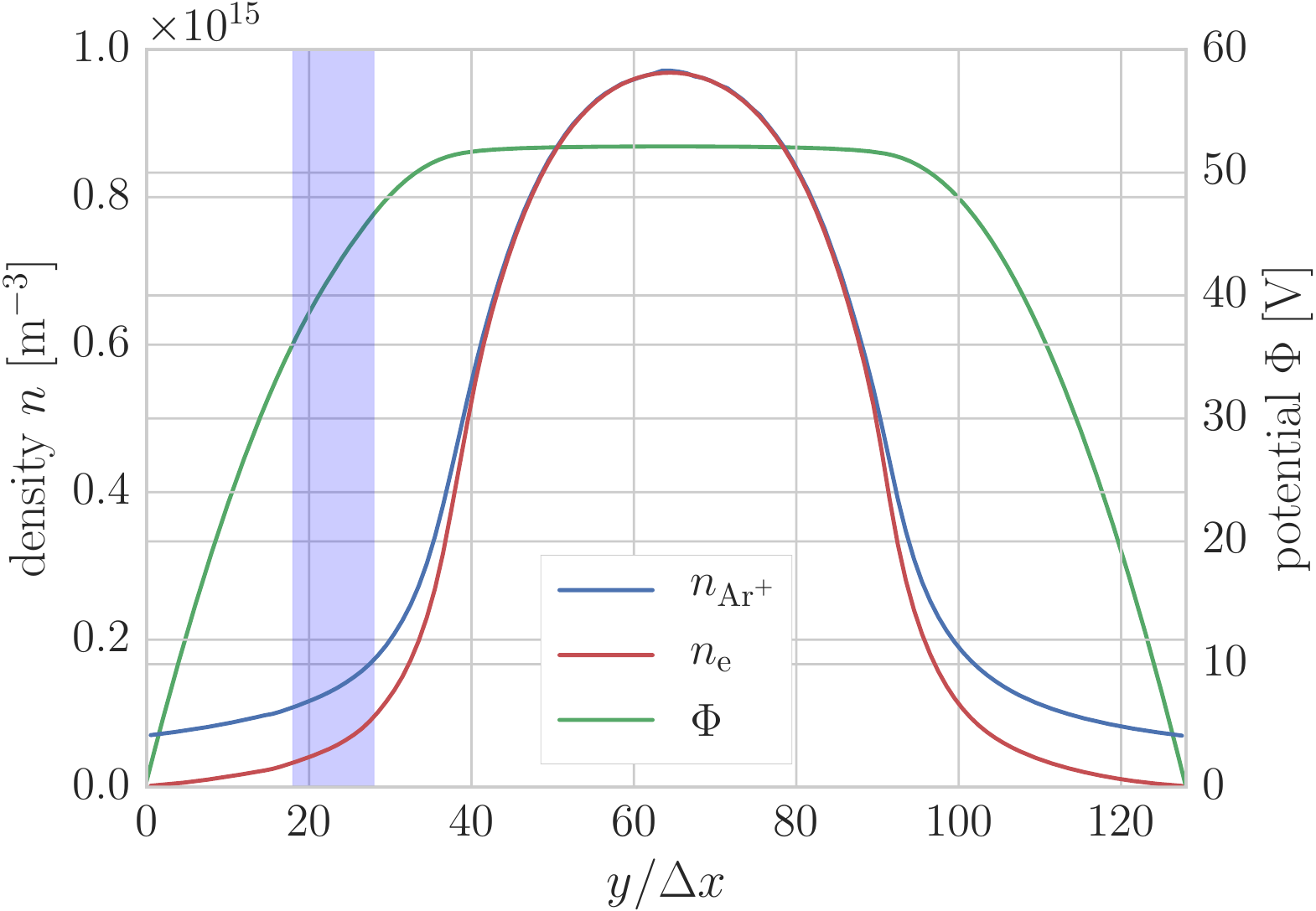}
\caption{(Color online)
  Average density and potential profile of the \chem{Ar} plasma along the discharge without dust particles.
  The position of the center of the dust cluster is given at $y = 24\Delta x \approx \SI{5.6}{mm}$ and 
  its extent is marked by a blueish vertical span. Averaging was performed over \num{250} rf periods and across 
  the lateral $x$ and $z$ axes.}
\label{Fig:DenPotBackground}
\end{figure}

Plasma properties of the discharge, such as densities and velocity distributions, determine the charging process of the individual dust particles inside the dust cluster.
The charges of the dust particles themselves determine the electrostatic potential and electric field within and outside of the dust cluster, that in turn influences the local plasma properties, i.e.\ the presence of a dust cluster will modify the local discharge characteristics.
Therefore, the study of the discharge characteristics is necessary to gain insights into the charging processes.

As a starting point we first simulate the argon plasma background and analyse its plasma characteristics without any dust particles included.
In Fig.~\ref{Fig:DenPotBackground} the computed electron and argon ion densities as well as the plasma potential in the discharge between the electrodes are shown.
After 500 rf periods the quasi-stationary state is analysed.
The data are time averaged over 250 rf periods and space averaged along $x$ and $z$ axis. 
The electron density equals the ion density in the bulk of the discharge, so that quasineutrality is satisfied and the plasma potential is constant.
The bulk plasma has a density $n_e = n_\chem{Ar^+} \approx \SI{9.7e14}{\meter^{-3}}$ and a potential $\Phi \approx \SI{52.1}{\volt}$.
Towards the electrodes the densities and potential decrease.
In this sheath region the ion density exceeds the electron density and the plasma potential drops towards zero.
At the electrodes we find a finite ion density of about $n_\chem{Ar^+} \approx \SI{7e13}{\meter^{-3}}$ and in comparison a vanishing electron density.
The sheath width can be derived from the plot to be about $\num{40}{\Delta x} \approx \SI{9.4}{\milli\meter}$.
Within the lower sheath (left half in Fig.\ \ref{Fig:DenPotBackground}) the shadowed region between $y=\numrange{18}{30}{\Delta x} \approx \SIrange{4.2}{7}{\milli\meter}$ marks the position and roughly the axial extent of the dust cluster.
A summary of the background plasma densities and potential in this region is given in Table \ref{Tab:DensPot} below.

\begin{table}[bth]
\caption{\label{Tab:DensPot}
  Overview of electron and argon ion densities and the electrostatic potential at different axial positions in the discharge: in the bulk, at the upper, center and lower position of the dust cluster and at the electrode.}
\begin{tabular}{l @{\quad} c @{\quad} c @{\quad} r}
\toprule
Position & Electron density & Argon ion density & Potential \\
$y/\Delta x$ & $n_\mrm{e}\ [\si{\meter^{-3}}]$ & $n_\chem{Ar^+}\ [\si{\meter^{-3}}]$ & $\Phi\ [\si{\volt}]$ \\
\colrule
bulk \\
64 & \num{9.7e14} & \num{9.7e14} & \num{52.1} \\
\colrule
dust \\
30 & \num{1.3e14} & \num{2.1e14} & \num{48.2} \\
24 & \num{6.4e13} & \num{1.4e14} & \num{43.0} \\
18 & \num{3.5e13} & \num{1.1e14} & \num{36.1} \\
\colrule
electrode \\
0 & \num{2e12} & \num{7.1e13} & \num{0.5} \\
\botrule
\end{tabular}
\end{table}

\begin{figure}[tb]
\includegraphics[width=0.99\linewidth]{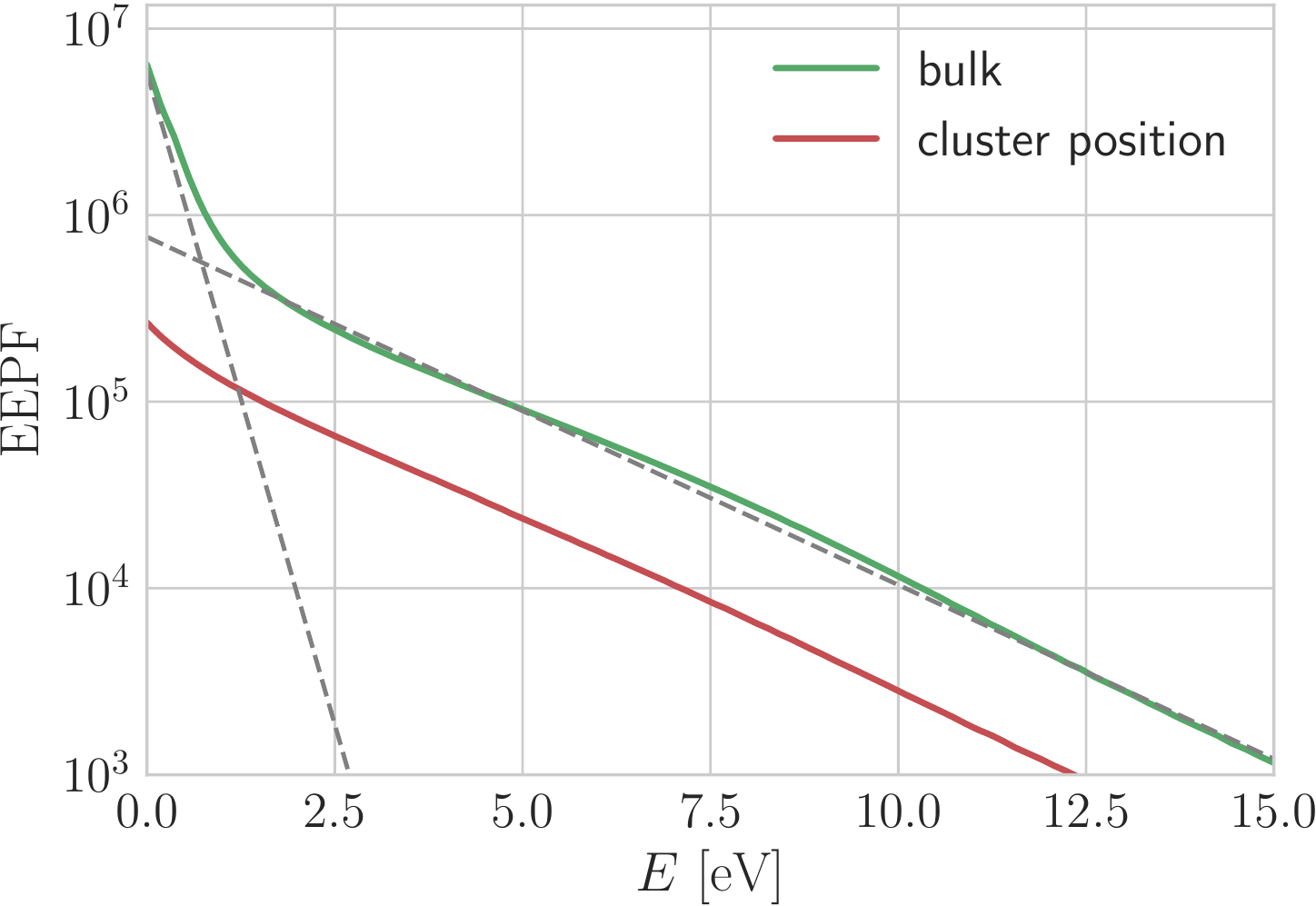}
\caption{\label{Fig:EEPFBackground}(Color online)
  Average electron energy probability function (EEPF) in the bulk of the discharge (green line) and at the proposed position of the dust cluster (red line). 
  Dashed lines are Maxwellian fit functions to the data.
  The data was acquired without dust particles in the simulation.
  Averaging was performed over \num{250} rf periods and across the lateral $x$ and $z$ axes.}
\end{figure}

Figure~\ref{Fig:EEPFBackground} shows the energy probability function of the plasma electrons (EEPF) at two different positions of the discharge.
For the center of the discharge (bulk) the data show that the electrons are represented by two Maxwellian distributions with different temperatures.
The effective temperatures have been obtained by fitting the data to two Maxwellians, shown as dashed lines in the plot.
The electrons in the plasma bulk have a temperature of $T_l = \SI{0.31}{\electronvolt}$ and $T_h = \SI{2.32}{\electronvolt}$ for the low- and high-temperature groups, respectively.
This feature is known from gases like argon that have a large Ramsauer effect and results from the stochastic heating of electrons due to the oscillating plasma sheath boundary in low pressure rf discharges.\cite{godyak_abnormally_1990}
Comparable temperatures for an experimental argon plasma with a higher pressure have been found by Godyak~et~al.\cite{godyak_abnormally_1990}.
The EEPF within the plasma sheath at the proposed center position of the dust cluster (cf.\ Fig.~\ref{Fig:DenPotBackground}) is shifted downwards because of the lower electron density there and merely shows the same slope as the high temperature electrons in the bulk.

%------------------------------------------------------------------------------ 
% Results - Charging
%------------------------------------------------------------------------------
\subsection{Charging of the dust cluster}
\label{Sec:Charging}

\begin{figure}[tb]
\includegraphics[width=0.99\linewidth]{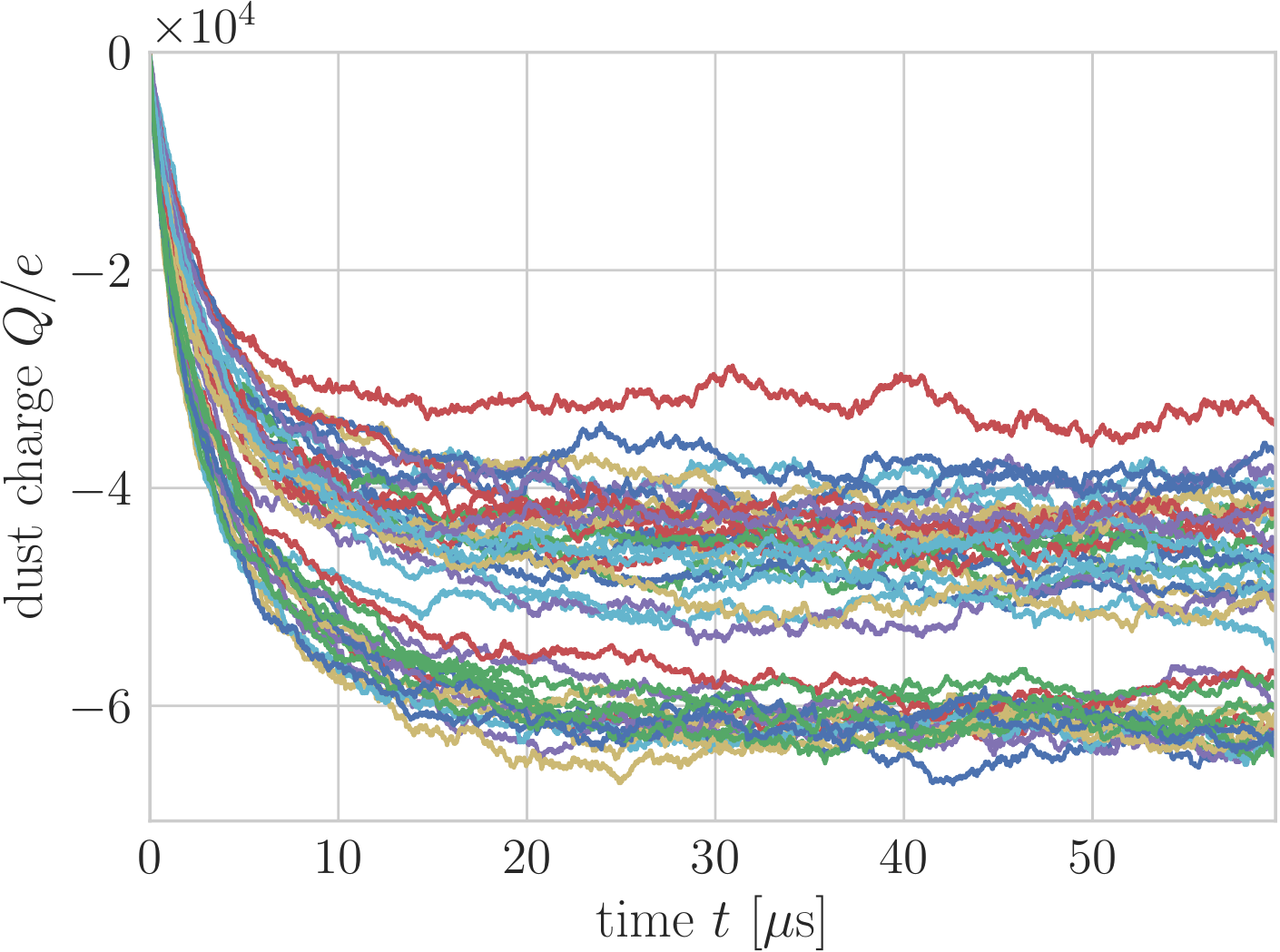}
\caption{\label{Fig:DustCharging}(Color online)
  Time evolution of the charge for every dust particle of the 3D dust cluster.
  Equilibration time is around \SI{30}{\micro\second}.}
\end{figure}

\begin{figure*}[tb]
\begin{minipage}[b]{0.49\columnwidth}
\includegraphics[width=0.99\linewidth]{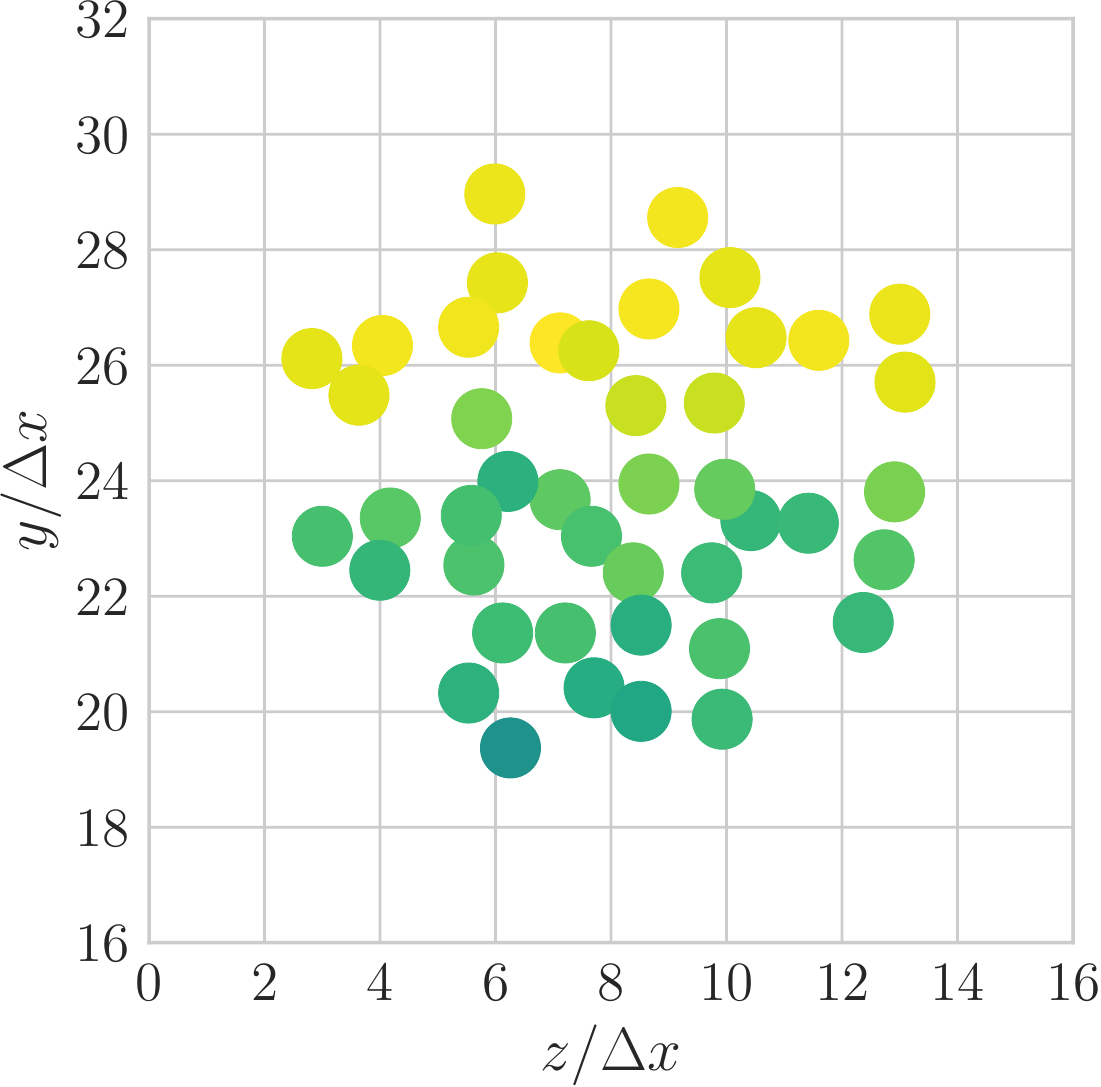}
\includegraphics[width=0.99\linewidth]{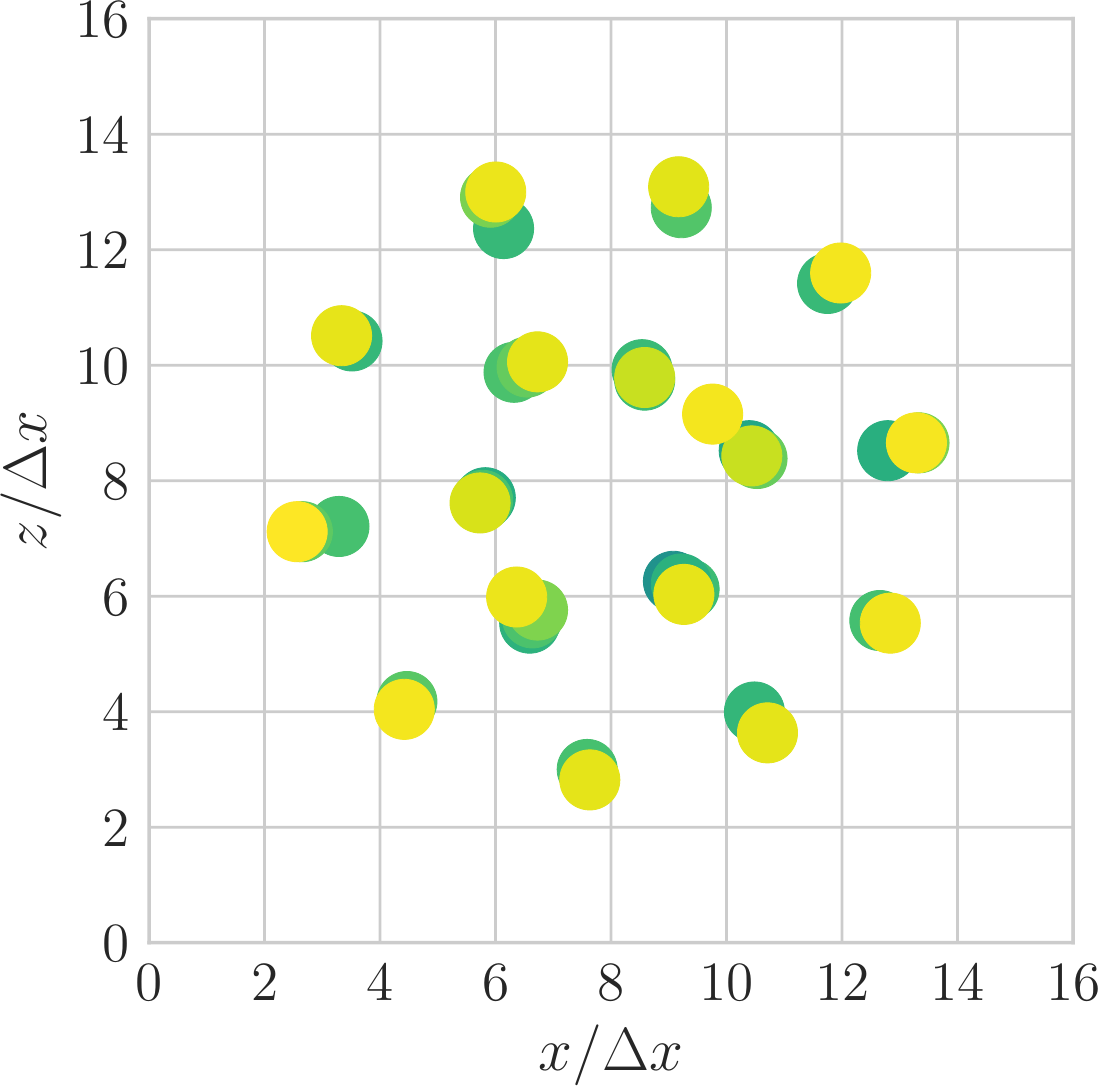}
\end{minipage}
\includegraphics[width=0.49\linewidth]{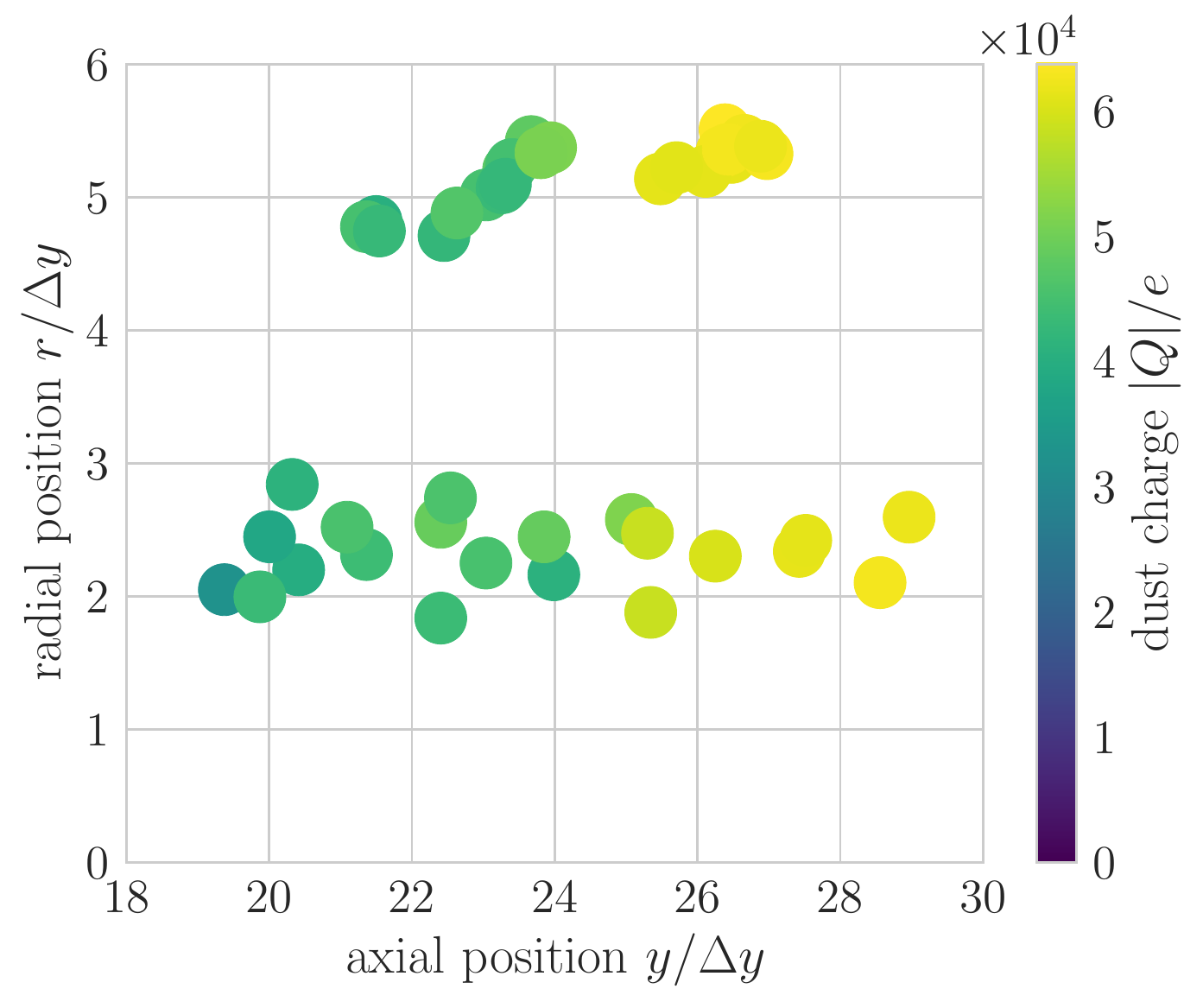}
\caption{(Color online)
  Average dust charge for dust particles in the 3D dust cluster from different perspectives and projections.
  The absolut charge value of a dust particle is color-coded. On the left-hand side, similar to Fig.\ \ref{Fig:Dust}, the dust particle positions are shown from different perspectives: side view (upper panel) and top view (lower panel).
  On the right-hand side the radial position of the dust particles from the center of the whole dust cluster $r = \sqrt{x^2+z^2}$ is plotted against its axial position in the discharge $y$.
  The plasma flow through the dust cluster streams towards $y=0$ and $\Delta y = \Delta x = 234.28\, \mu m$.}
\label{Fig:DustCharge}
\end{figure*}

As the plasma discharge is in a dynamic equilibrium the next step is to immerse the dust cluster, as described in Sec.\ \ref{Sec:Setup}, into the simulation.
The dust cluster has a rigid structure, that is, the positions of the dust particles in relation to each other are fixed.
The cluster and its orientation resembles the original experiment.
The center of mass of the cluster is pinned at $y = \SI{5.6}{\mm}$ above the lower electrode.
The initial charge of each dust particle is zero.
The location of the dust cloud is within the sheath of the rf discharge.
As a consequence, strong asymmetric flows towards the electrode within the sheath, modulated by the rf dynamics, act on the dust particles.
The average ion flow at the dust positions is supersonic and varies between 1.1 and 2.1 Mach.

The time evolution of the charge for every dust particle of the dust cluster is depicted in Fig.~\ref{Fig:DustCharging}.
The dust particles quickly gain a negative charge but the absolute values differ significantly as can be seen by the spread of charges.
The ratio between maximum and minimum absolut value is about $\num{2}$.
After around $t_\mrm{eq} = \SI{30}{\micro\second}$ the charges saturate and a quasi-stationary equilibrium with charges between \SIrange{-30000}{-62000}{e} is reached.
Nonetheless the values of charge slightly fluctuate.
Fluctuations on the small scale are due to the discrete nature of our simulation.
Each time a super-particle (ion or electron) is absorbed by a dust particle the charge increases or decreases by a multiple of the elementary charge according to the super-particle ratio.
On a larger scale at the order of the rf cycle the repeated flooding and depletion of the cluster environment with electrons in every rf cycle leads also to charge fluctuations.
Based on the simulation results for the charging of a single dust grain the fluctuations should be only on the order of a few percent~\cite{melzer_phase-resolved_2011} and thus hardly visible on the scale of Fig.~\ref{Fig:DustCharging}. 
In reality the charges could also fluctuate provided the material-dependent desorption time for electrons~\cite{HBF11} collected by the particle is shorter than the rf cycle. 
In the simulation this effect is not included because particles are assumed to be perfect absorbers.
In the next subsection we visualize the flooding and depletion of the cluster environment with electrons by tracking, at quasi-stationarity, the electron flux over one rf period. 

%------------------------------------------------------------------------------ 
% Results - Charge distribution
%------------------------------------------------------------------------------

\begin{figure}[tb]
\includegraphics[width=0.99\linewidth]{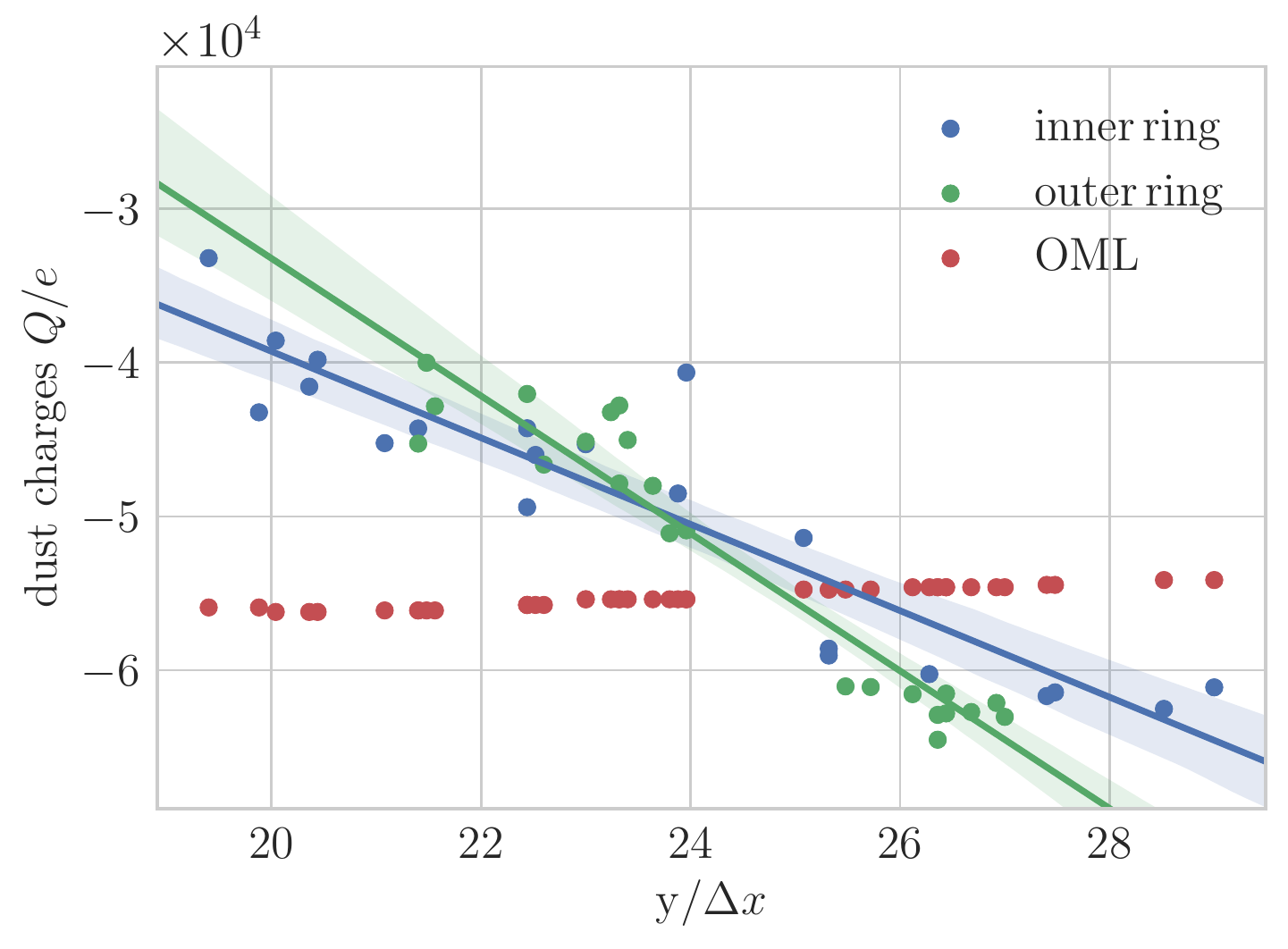}
\caption{(Color online)
  Average charge of the grains in the 3D dust cluster along their axial positions $y$ in the discharge
  which is measured in units of $\Delta x = 234.28\, \mu m$.
  Linear regressions with $95\%$ confidence interval are also shown.
}
\label{Fig:DustChargeAxial}
\end{figure}

\begin{figure*}[tb]
\includegraphics[width=0.99\linewidth]{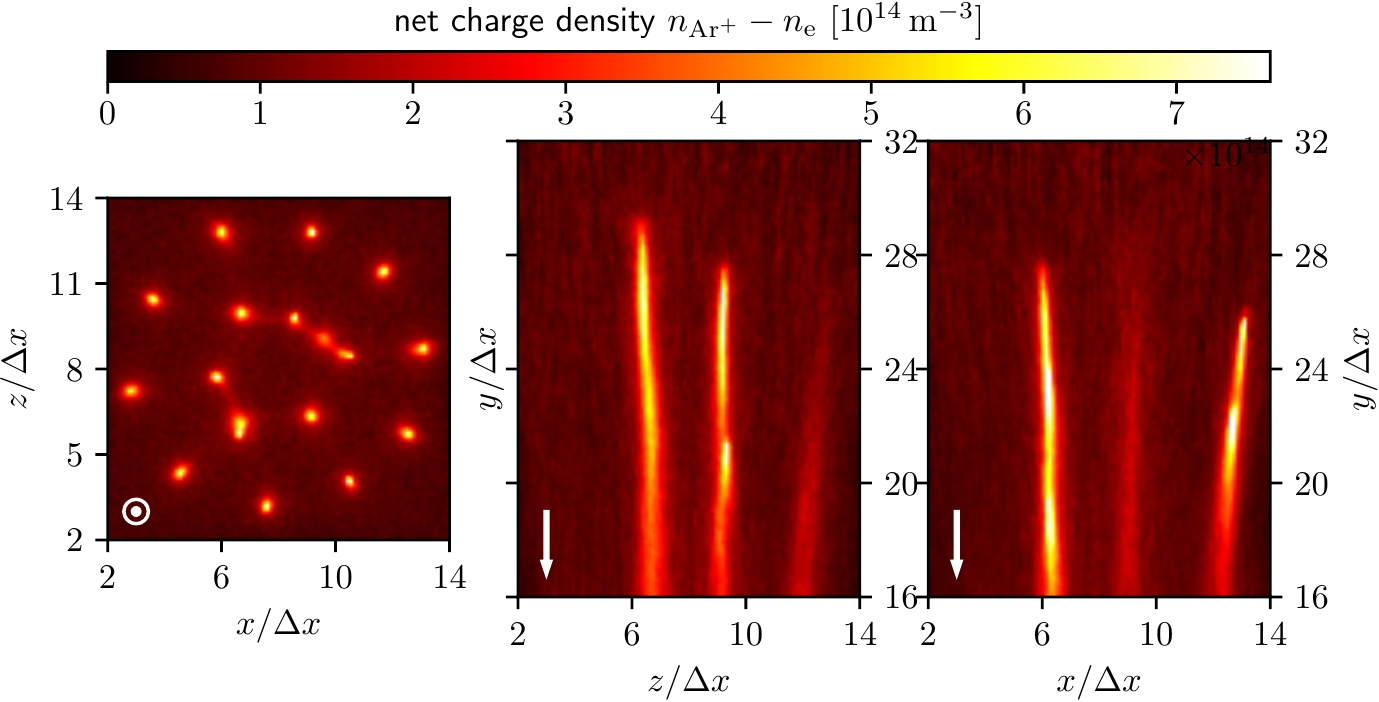}
\caption{(Color online)
    Net charge density $n = (n_\chem{Ar^+}-n_\chem{e})$ shown for $xz$ (left), $yz$ (middle) and $xy$ (right) cut planes through the dust cluster.
    Lighter colors represent higher densities. The unit of length is $\Delta x = 234.28\, \mu m$.
    The arrows indicate the net flow direction of the plasma ions.}
\label{Fig:MDdensitywakes}
\end{figure*}

Figure~\ref{Fig:DustCharge} allows a view on the spatial arrangement of the quasi-stationary dust grain charges.
From the top panel it can be seen that the upper-most dust particles accumulate most of the charge.
Going downstream the absolute value of charge of the dust grains decreases continuously.
Plotting the radial distance from the symmetry axis against the axial position of the dust particles no radial dependence of the charge can be found.
The inner and outer ring of the cluster each show a continuous decrease of charge toward the lower electrode.
Therefore, the only significant functional dependence for the accumulated charge is the variation with the axial position of the dust grains as shown in Fig.~\ref{Fig:DustChargeAxial}. 

To estimate how strong the charging of a particular grain is influenced by the others we compare its simulated charge with the charge the particle would have obtained in the hypothetical case that no other grains were around.
The hypothetical charges, shown in Fig.~\ref{Fig:DustChargeAxial} by the red symbols, are obtained by orbital-motion-limited (OML) balancing of electron and ion fluxes, including the ion flux due to charge-exchange collisions and taking the local electron and ion densities as well as the local electron and ion temperatures into account (as it was done in another context in~\cite{BFK09}).
Clearly, the hypothetical charges do not depend on the axial position of the grains whereas the simulated charges show a linear dependence over the axial position, with a slope depending on whether the particles belong to the inner or the outer ring. 
%The fit indicates that there is a jump and hence a deviation from the linear dust charge dependence.
Particles at the upper side of the cluster, facing the bulk plasma, collect charges as if the others were not present.
The closer a particle is to the powered electrode, the stronger is the discrepancy between simulated and hypothetical charge indicating mutual influencing of the charging process. 
To understand this process and hence the charge distribution within the cluster we now take a look at the electron and ion fluxes prevailing in its vicinity. 

%------------------------------------------------------------------------------ 
% Results - Plasma environment
%------------------------------------------------------------------------------

\subsection{Plasma modification by the dust cluster}
\label{Sec:PlasmaEnv}

In Fig.~\ref{Fig:MDdensitywakes} the plasma charge density for three orthogonal cut planes through the plasma enviroment of the dust cluster is given for the quasi-stationary situation after the fast charging phase within the first rf period. 
All quantities are time-averaged over 250 rf periods, both for phase-resolved and phase-averaged diagnostics.
For the diagnostics within the MD region a $\num{200 x 200 x 200}$ cubical grid is used instead of the $\num{16 x 16 x 16}$ cubical PIC cells.
This allows to resolve the potential, densities, velocities and fluxes averaged over the distribution functions in the MD region with this fine grid.

For each visible dust particle long reaching wake fields can be seen behind them, indicated by an increased positive charge density, i.e.\ an excess of ions in that space region.
In the middle panel multiple dust particles align within the wake field of the upper particles and form linear chains.
The diameter of these individual particle wakes is roughly $d \approx \num{0.5} \Delta x \approx \SI{117}{\micro\meter}$, so they are around 6 times larger than the dust particle diameter itself.
Also an overall focusing effect of these wake fields towards the symmetry axis of the cluster is evident.
It is caused by the attraction of the ions to the large negative charge of the dust cluster as a whole.

The self-consistent kinetic description of the rf discharge interacting with the ensemble of dust particles allows us to visualize also the plasma flows around and inside the dust cluster.
Thereby, we can make a first step towards an understanding of the interaction of a collection of solid bodies with a time-dependent plasma, in contrast to the interaction with a stationary plasma usually considered. 
\begin{figure*}[tb]
\includegraphics[width=0.99\linewidth]{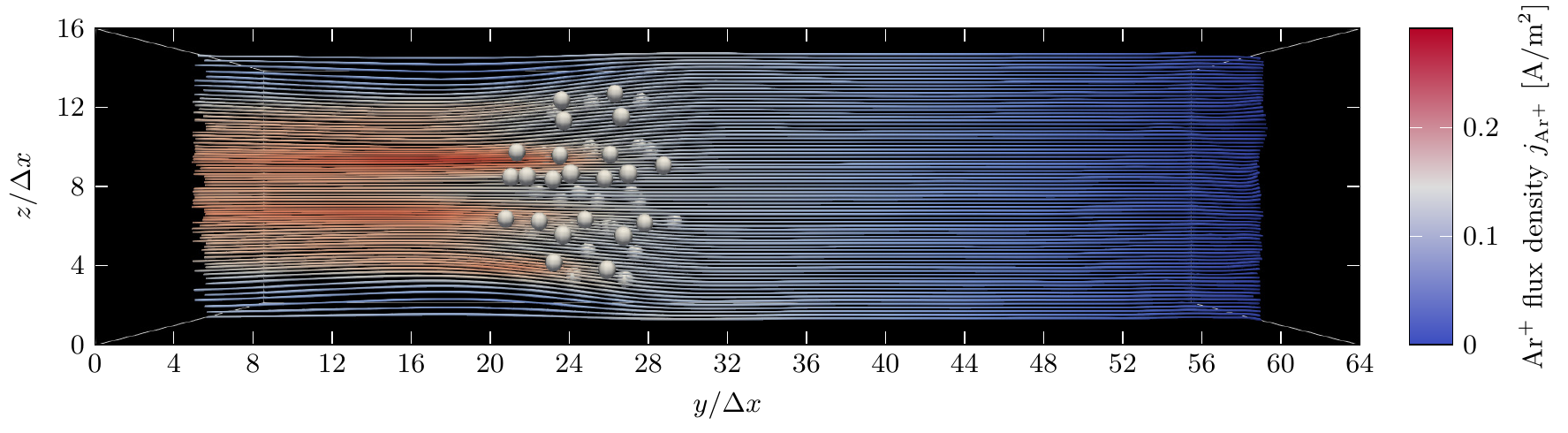}
\caption{(Color online)
    Streamlines of the particle flux for \chem{Ar^+} ions sampled over PIC cells with extension 
    $\Delta x = 234.28\, \mu m$ in all three spatial directions and time-averaged over 250 rf periods.
The direction of the flux is from the plasma bulk (right) to the electrode (left).}
\label{Fig:PIC_ion_flow}
\end{figure*}
We use the modeling results to study first the flow effects on ions and electrons on the PIC scale. 
Due to their large mass ions do not show a significant rf modulation effect, because they react on a time-scale longer than the rf period.
Therefore, for ions an additional time-averaging over the rf period can be done, whereas for electrons the phase-resolved particle fluxes are calculated, because they are strongly influenced by the rf voltage modulation.
In the following tracer visualization of fluxes using ParaView \cite{Paraview} is done.
The ion fluxes in Fig.~\ref{Fig:PIC_ion_flow} show the focusing effect on the ion flow discussed before.
The ions experience the ensemble of dust grains as an additional negative charge, that is, as a reservoir of (bound) electrons which would not be there without dust.
This attracts the overall flow of ions towards the dust ensemble with a global streaming in the direction towards the electrode.
As a result of this dynamic effect, arising from the self-consistent description of the spatio-temporal dynamics of the rf plasma and the charge accumulated by the dust grains, a modification of the surrounding plasma on the spatial scale of the dust ensemble is observed. 

\begin{figure*}
\includegraphics[width=0.99\linewidth]{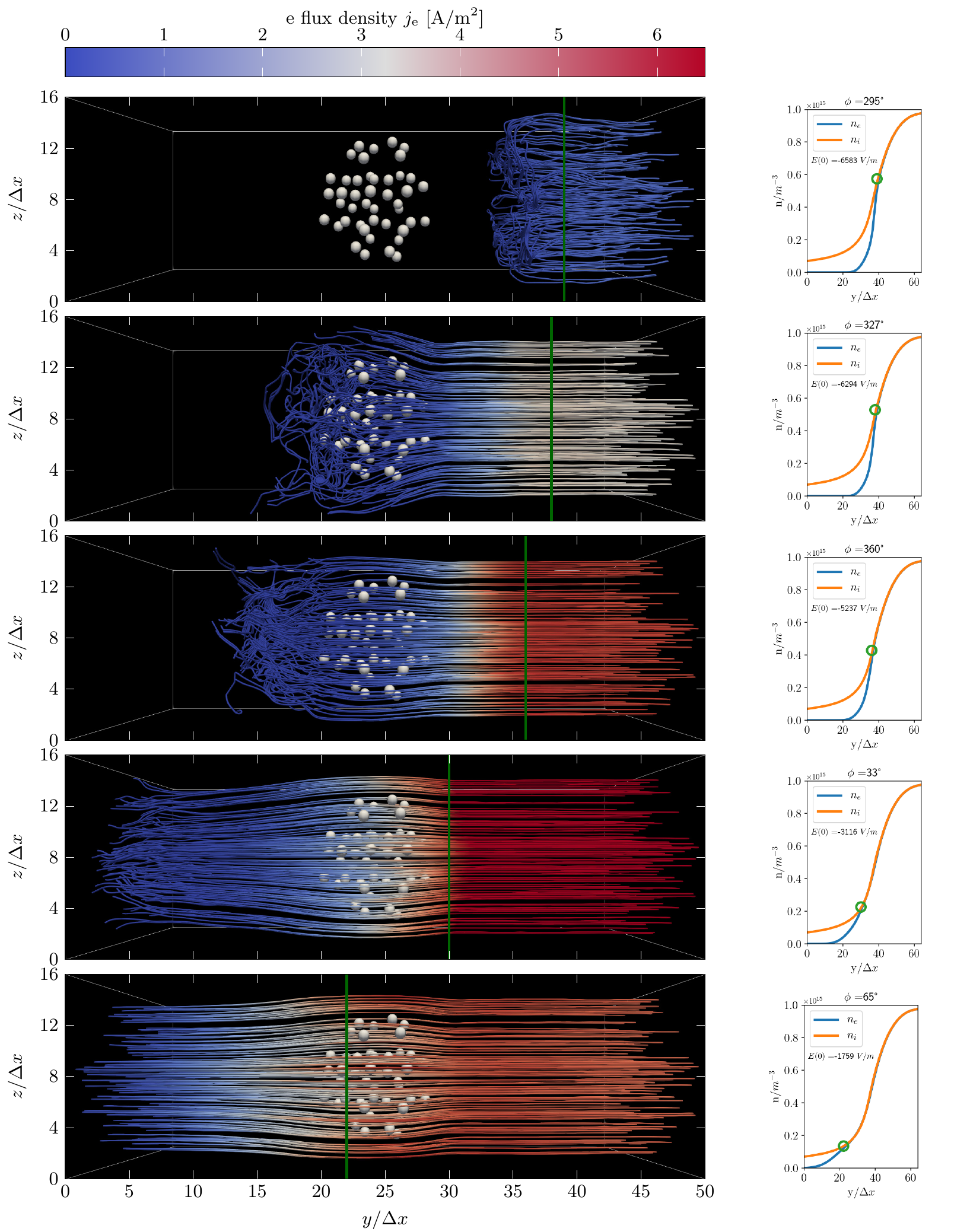}
\caption{(Color online)
  Left panel: Time-resolved electron flux sampled over PIC cells with spatial extension $\Delta x = 234.28\, \mu m$ for selected 
  phases within the rf period. The vertical line indicates the position of the moving rf sheath entrance. Right panel: Electron 
  and ion densities at the corresponding rf phases without dust. The green circle corresponds to the vertical line on the left 
  panel where quasi-neutrality breaks. The unit of length is again $\Delta x = 234.28\, \mu m$.}
\label{Fig:GlobalPlasma}
\end{figure*}

\begin{figure*}
\includegraphics[width=0.95\linewidth]{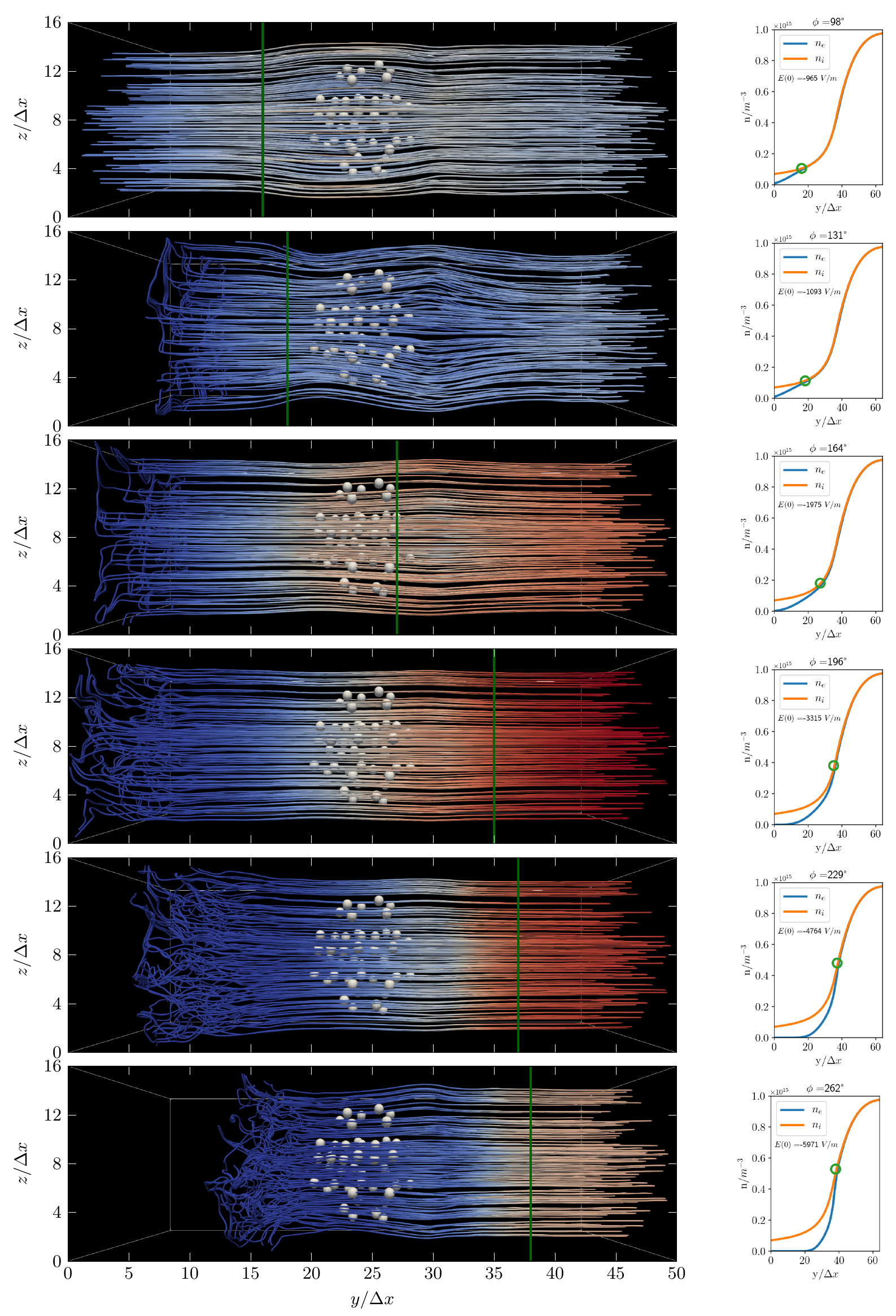}
\caption{(Color online)
Figure \ref{Fig:GlobalPlasma} continued.}
\label{Fig:GlobalPlasma2}
\end{figure*}

\begin{figure*}[tb]
\includegraphics[width=0.95\linewidth]{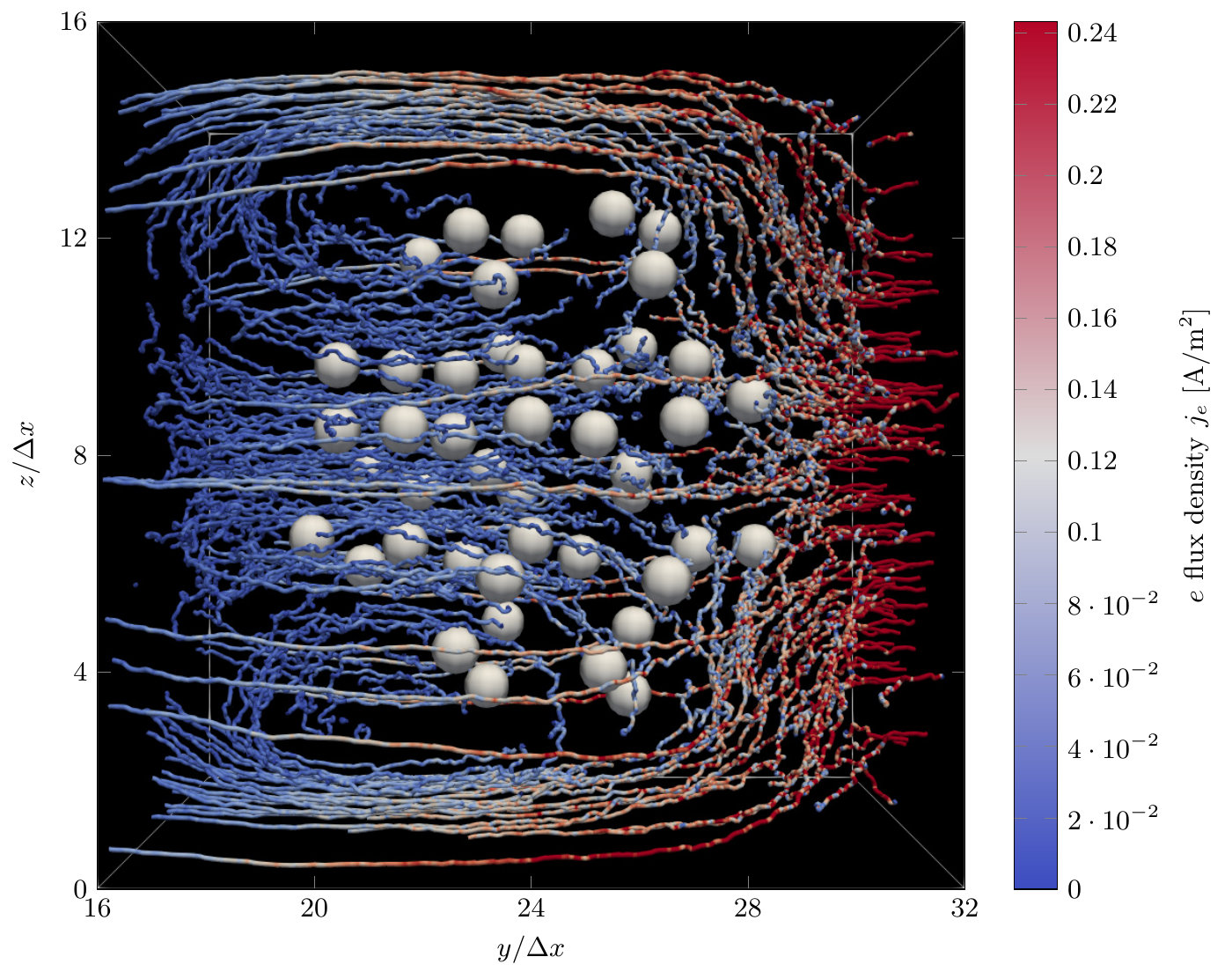}
\caption{(Color online)
Electron flux within the MD region sampled over MD cells with spatial extension $0.08 \Delta x$, where 
$\Delta x = 234.28\,\mu m$, and time-averaged over 250 rf periods.}
\label{Fig:MDflux}
\end{figure*}

\begin{figure*}[tb]
\includegraphics[width=0.95\linewidth]{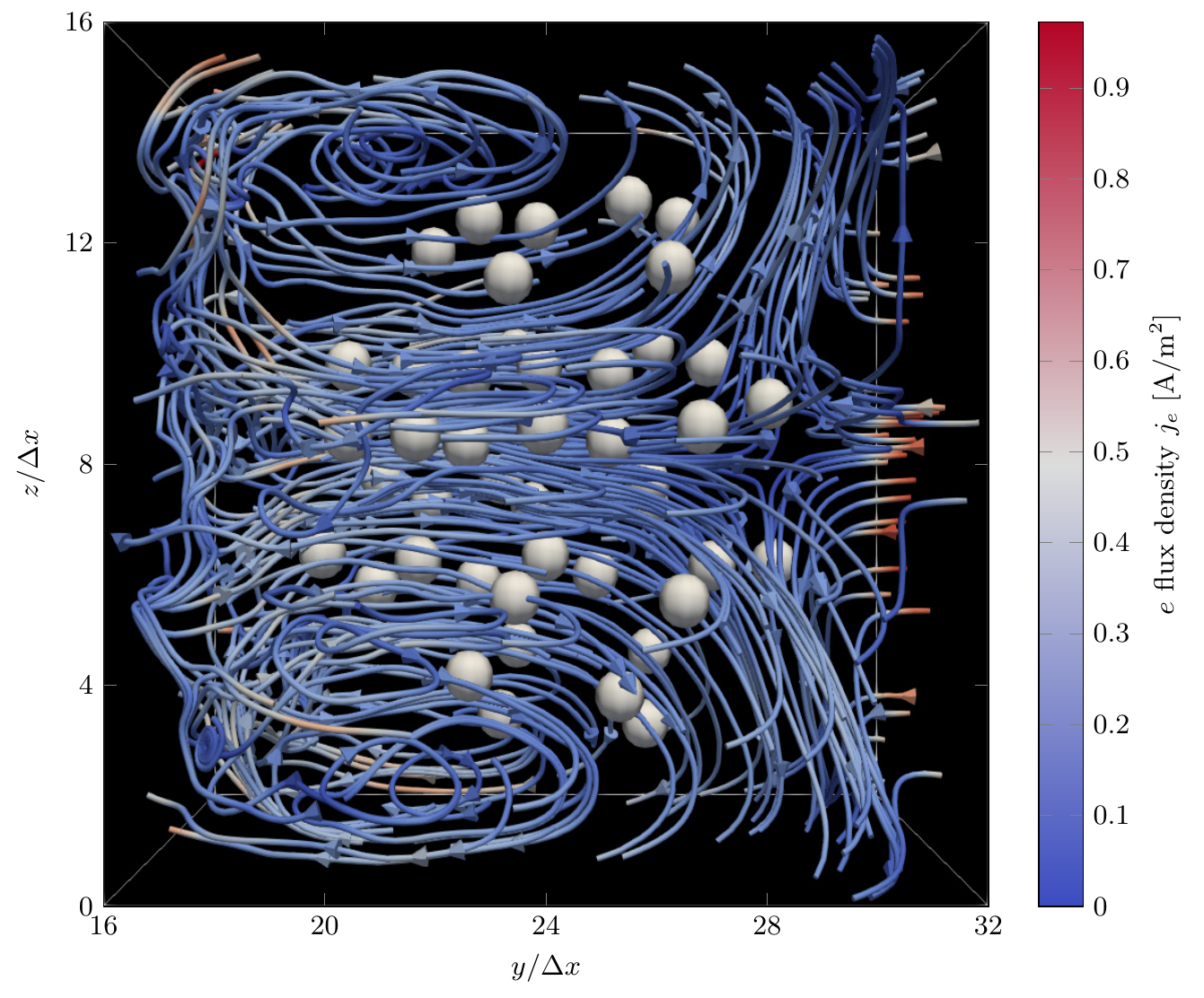}
\caption{(Color online)
Electron flux within the MD region sampled over PIC cells with spatial extension $\Delta x = 234.28\, \mu m$, that is, over cells 
which are 12.5 times larger in each spatial direction than the cells used for obtaining the data shown in Fig.~\ref{Fig:MDflux}. 
The flux is again time-averaged over 250 rf periods.}
\label{Fig:PICMDflux}
\end{figure*}

The electron dynamics exhibits much more complex features due to the small mass of the electrons.
Individual dust particles in the sheath act as rocks in the surf, because the mobile electrons pass them twice during one rf period 
as a kind of density front according to the rf modulation, whereas the plasma ions are nearly immobile over one rf period.
The electron front passes an individual dust particle once towards its way to the electrode and once towards its way to the bulk plasma.
This dynamics leads already for a single dust particle in the sheath to phase-dependent dust charges~\cite{melzer_phase-resolved_2011}.
In the present study an ensemble of dust particles interacts with the rf plasma.  Figures.~\ref{Fig:GlobalPlasma} and ~\ref{Fig:GlobalPlasma2})
show the phase-resolved electron flux streamlines for different rf phases with the position of the moving rf sheath entrance indicated
by the green vertical lines. The rf modulation first drives the front towards and through the dust ensemble and then pushes it back again.
The strongest interaction between the dust and plasma takes place in the phases, when the electron front intersects the dust sphere.
Notice, at $\Phi = 98\degree$ an increase of electron flux both in front and in the back of the dust cloud is observed.
Here, the electric field at the electrode has its smallest value. This bi-modal build-up of density, occurring when the cluster is surrounded 
temporarily by a quasi-neutral plasma, is created by the blocking of the rf driven electron flow by the porous dust cloud. 
%Such an effect is very similar to fluid dynamics or aerodynamics of flows around objects. 
The rather turbulent trajectories close to the electrode are a consequence of the quite small electron density in this region. 
Therefore, the flow field is no longer statistically averaged, but a result from individual particle trajectories, which then show up 
also in the visualisation.

Further insights into the plasma dynamics are obtained by resolving on the MD scale the electron flux inside the dust ensemble. It 
is shown in Fig.~\ref{Fig:MDflux}. Here, the flow fields are averaged over the individual MD cells and not over the much larger 
PIC cells. We present data only for rf averaged flow fields (averaged as before over 250 rf periods) because the statistics at a
given instant of time of the time-resolved flows is rather noisy. The penetration of electrons into the dust cloud is clearly visible 
which hence acts on this scale as expected like a finite size porous medium. Notice also that the reservoir of background electrons is 
apparently not sufficient for the charging of all of the dust particles. Electrons are not only sucked in from the major laminar flow 
directed towards the electrode, but also radially from the side. For comparison we plot in Fig.~\ref{Fig:PICMDflux} also the electron 
flux in the vicinity of the dust cluster obtained by averaging over PIC cells. In both cases we find electrons flowing partially through 
the dust cloud and partially around it giving rise to convective cell patterns. On the PIC scale, however, due to the averaging over the 
larger cell volumes, the dynamics is smoother. 

A technical remark is perhaps in order at this point. For Figs.~\ref{Fig:GlobalPlasma} 
and ~\ref{Fig:GlobalPlasma2} the sampling of the initial positions for the ParaView flow markers was done on a sphere on the right 
hand side of the figures. The markers visualize thus the flow around the dust and have practically no chance penetrating into the dust 
structure. For Figs.~\ref{Fig:MDflux} and~\ref{Fig:PICMDflux}, on the other hand, the sampling of initial positions for the markers 
is modified to resolve the full flow structures also inside the dust cloud. Here, two spheres are used, one at y=32 $\Delta x$ and one at 
y=16 $\Delta x$, such that the surface of the sphere intersects the cluster to visualize also the internal dynamics. In the phase-resolved 
electron dynamics (not shown because of the pure statistics) one identifies then flux-reversal zones for some parts of the rf pulse. 
Despite the negative charge of the grains, which would one expect to push electrons away from the cloud, electrons are thus also flowing 
towards the grains. The time-averaged convection patterns in Figs.~\ref{Fig:MDflux} and~\ref{Fig:PICMDflux} are a result of this 
flux reversal. This is consistent with the dust charge distribution in the cloud as discussed before. Grains downstream the dominating 
laminar plasma flow have to collect electrons also radially because of the depletion of electrons inside the cluster. The competition 
for electrons thus favors also convective flows.  
%In combination with the flow-reversal the competition for electrons drives convection patterns on both the MD and PIC scale. 

Finally we comment on the ion flux which 
we do not show on the  MD scale. It suffers no reversal, stays hence laminar also on the MD scale, because the inertia of the ions prevents 
them to react on such short (time) scales to the charge of the dust particles. The ions remain in their supersonic laminar flow field 
required to stabilise the plasma sheath hosting the dust ensemble. Shadowing and focusing effects appear on the PIC scale, that is,  
on the scale of the Debye length, as shown in Fig.~\ref{Fig:PIC_ion_flow}, and more pronounced close to the dust grains 
as discussed in a previous work~\cite{IMM10}.

%------------------------------------------------------------------------------ 
% Conclusion
%------------------------------------------------------------------------------
\section{Conclusions}
\label{Sec:Conclusions}

We have carried out a three-dimensional plasma simulation using a PIC-MCC/PPPM code, to compute the dust charges of a rigid three-dimensional dust cluster immersed in the sheath of an argon rf gas discharge.
The geometry of the cluster, its position in the sheath, and the plasma parameters used for the simulation are typical for experimental set-ups. 
We expect therefore the simulation to reproduce rather closely the self-consistent dynamic electric response of an rf plasma to an ensemble of dust particles and vice versa as it occurs in actual experiments.
The mechanical response of the dust cloud to the plasma, that is, the adjustment of the grain positions to the electric fields acting on them, is not considered.
It occurs on a much longer time scale outside the scope of this investigation and would have in addition required to include also a lateral confinement of the grains preventing them to arrange in a two-dimensional layer.

As far as the charge of the dust particles is concerned we found the particles closest to the bulk plasma to carry the largest charge which in fact is rather close to the orbital-motion limited charge at that position in the plasma corrected by charge-exchange scattering and calculated with a directed ion flux.
Along the axial direction the charge of the grains diminishes towards the electrode indicating a strong competition of the downstream grains for electrons from the background plasma as well as a significant deviation of the plasma flow inside the dust cluster from the flow hitting the upstream grains closest to the bulk plasma.
The spread of the grain charges is rather large with downstream grains carrying only half as many electrons as the upstream ones.  

Simulating the full dynamics of the electric response of the rf plasma to the ensemble of dust grains and vice versa we could visualise the spatio-temporal structure of plasma flows around and inside the dust cloud on the scale of the Debye length (PIC scale) as well as the scale of the dust potential (MD scale).
As expected, the ion flux is rather laminar, because of the inertia of the ions, showing only shadowing and focusing effects on the scale of the Debye length or below.
The electron flux however is more complex.
The magnitude of the electron flux in front and inside the dust cluster depends on the phase within the rf period with a bi-modal modulation when the cluster is temporarily surrounded by a quasi-neutral plasma.
Grains downstream the dominating laminar plasma flow have to collect electrons also radially, because of the depletion of electrons inside the cluster.
The need to supply enough electrons forces thus the electron flux to develop convective cell patterns on both the MD and PIC scale.

%------------------------------------------------------------------------------ 
% Acknowledgement
%------------------------------------------------------------------------------
\section*{Acknowledgement}

This work was supported by the Deutsche Forschungsgemeinschaft through the Transregional Collaborative Research Center SFB/TRR 24. 
We thank J.\ Schablinski, D.\ Block and F.\ Greiner for providing information about the experimental setup and A. Melzer for 
discussions and reading a first draft of the manuscript. Numerical calculations were performed on the \textit{iapetos} compute 
cluster at the Institute of Physics, University of Greifswald. 

%\bibliographystyle{apsrev4-1}
%\bibliography{DustyPlasma,ref}

%

\end{document}